# Multiperiodic magnetoplasmonic gratings fabricated by the pulse force nanolithography

Savelii V. Lutsenko[1,*], Mikhail A. Kozhaev[2,5], Olga V. Borovkova[1,2], Andrey N. Kalish[1,2,3], Alexei G. Temiryazev[4], Sarkis A. Dagesyan[1], Vladimir N. Berzhansky[5], Alexander N. Shaposhnikov[6], Alexei N. Kuzmichev[2], Vladimir I. Belotelov[1,2,5]

[1] *Faculty of Physics, Lomonosov Moscow State University, Leninskie Gory, Moscow 119991, Russia*
[2] *Russian Quantum Center, Skolkovskoe shosse 45, Moscow 121353, Russia*
[3] *NTI Center for Quantum Communications, National University of Science and Technology MISiS, Leninsky Prospekt 4, Moscow 119049, Russia*
[4] *Kotel'nikov Institute of Radioengineering and Electronics of RAS, Fryazino Branch, Vvedensky Sq. 1, Fryazino 141190, Russia*
[5] *V.I. Vernadsky Crimean Federal University, Vernadsky Ave. 4, Simferopol 295007, Russia*
[6] *Research Center for Functional Materials and Nanotechnologies, Institute of Physics and Technology, V.I. Vernadsky Crimean Federal University, Vernadsky Ave. 4, Simferopol 295007, Russia*
*\*Corresponding author: savlucenko@yandex.ru*

**We propose a novel technique for the magnetoplasmonic nanostructures fabrication based on the pulse force nanolithography method. It allows one to create the high-quality magnetoplasmonic nanostructures that have lower total losses than the gratings made by the electron beam lithography. The method provides control of the SPP excitation efficiency by varying the grating parameters such as the scratching depth or the number of the scratches in a single period. The quality of the plasmonic gratings was estimated by means of the transverse magneto-optical Kerr effect that is extremely sensitive to the finesse of a plasmonic structure. © 2021 Optical Society of America**

Nowadays, magneto-optical (MO) techniques for the control of light attract a significant interest [1]. Magnetic field allows manipulating such optical parameters as intensity and polarization at a high speed. Thus, MO effects are considered for telecommunication applications and are widely used for optical isolators, magnetic field sensors, biosensors, and data recording [2-8]. Some materials that do not reveal a strong intrinsic MO response are still rather attractive for these applications since MO effects in them can be enhanced by the excitation of the optical modes such as waveguide modes, localized and propagating surface plasmons [9-20]. The latter can be excited for example in the patterned magnetic heterostructures [14]. We can recall periodic subwavelength gratings of noble metals (Au, Ag) as well as the combined noble-magnetic gratings used for the enhancement of MO effects.

For the efficient excitation of the propagating surface plasmon polaritons (SPPs) it is necessary to fabricate the magnetoplasmonic nanostructures of proper geometry and size. Up to now several techniques have been applied for this purpose [21-26]. Usually the electron-beam lithography (EBL) method is used since it provides a high resolution up to 10 nm and versatile pattern formation [25]. In the direct writing EBL a small spot of the electron beam is written directly onto a resist coated substrate. The resulting nanostructured resist is used as a mask for subsequent layers applied to the sample. Another method is the nanoimprinting lithography with resolution pattern of 2 nm and a density range of $10^{12}$ dots/cm$^2$. However, it has some critical issues that need to be addressed for the further progress of this technology: its limitations in handling complex patterns with varied feature density, patterning over topographies and pattern alignment, to name the most important ones [21].

Also, a high-quality structure can be produced by etching a material using focused ion beams. It must be noticed that the mean-square surface roughness is around 0.3 nm. But it is worth noting the uncontrolled effect that ions have on the material, and therefore the properties of the material may be distorted. In addition to this, the production time is directly related to the energy of the ions. [26].

Another noteworthy technique is a vacuum annealing for the nanoisland structures formation [23,24]. The size and location of the 'islands' depend on the time and temperature of the vacuum annealing. Thereby it is possible to create an interaction between localized and propagating plasmon modes. However, this approach doesn't provide the fine control of the resulting parameters. In addition, the annealing speed depends on the applied temperature, for fast and efficient creation of structures, it is required to maintain temperatures of the order of 1000 degrees.

In this Letter, we propose a new method for fabrication of the magnetoplasmonic structures based on a pulse force nanolithography method (PFNL) [22]. It employs a rapid indentation of the sharp tip from one point to another with a help of a single crystal diamond tip. This approach provides a deep nanopatterning with high resolution in solid materials, like metals. PFNL method allows us to fine-tune the geometry of the structure providing an excitation of SPP modes with higher quality. We address the structures fabricated by the PFNL method in terms of the magneto-optical properties. The PFNL structures are compared with the ones made by EBL method.

We consider the sample composed of a dielectric magnetic layer covered with metal gratings that supports an excitation of the propagating SPP modes and waveguide modes in the magnetic layer. A 4-μm-thick magnetic dielectric film of bismuth-substituted iron garnet (BIG, $(YBi)_3(FeAlSc)_5O_{12}$) was grown on the gadolinium gallium garnet ($Gd_3Ga_5O_{12}$) substrate by the liquid phase epitaxy method. Then, a gold layer of $d$ = 40 nm thickness was deposited on top of it by the magnetron sputtering method. After that, the gold layer was scratched by PFNL method (see Fig. 1 (a)) [22]. Gratings of 131 blocks with a period $P$ = 690 nm were created in the sample. Each block is an array of several parallel grooves about 15 nm wide surrounded by hills of squeezed gold made by means of the nanolithography. The addressed structures are characterized by the distance between the lines in one block, $p$, and the number of lines in a single block, $n$. Two gratings 100×100 μm in size were made with different groove densities in a block. Parameters of the first structure (denoted here and further as "A") are $p$ = 75 nm, $n$ = 5, and the second structure (denoted as "B") are $p$ = 100 nm, $n$ = 4. In Fig. 1 one can find images of the structure "B" made by atomic force microscopy (b). The block sizes in both structures were approximately the same.

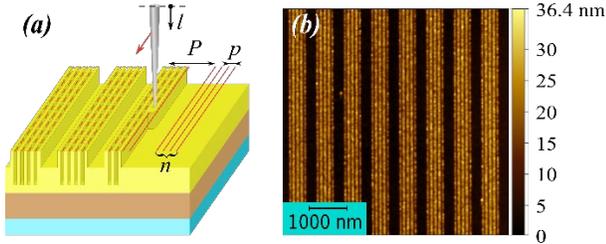

Fig. 1. (a) Scheme of the sample nanostructured by the PFNL technique. (b) Atomic force microscopy images of structure "B".

Besides that, the EBL structure (denoted as "E") (Fig. 2 (b)) was made on the same sample for the comparison with the PFNL structure (Fig. 2 (a)). This "E" structure was fabricated as follows. First, a film of a sensitive polymer (electronic resist) was applied on top of the substrate. In our case, it was polymethyl methacrylate (950 PMMA A2) deposited by centrifugation. Gold film was formed by the magnetron sputtering in a direct current discharge in argon at a pressure of 5×10$^{-3}$ mbar. For etching a gold film through a polymer mask, a conventional physical sputtering of the material in an argon discharge was used. Then, the remains of the polymer mask were removed using an oxygen plasma discharge. The period of this grating is $P$ = 690 nm, the same as of the PFNL structure. The width of the air gap is 115 nm, for the compared PFNL structures it's 290 nm and 315 nm, respectively.

The transverse magneto-optical Kerr effect (TMOKE) was measured to study the properties of the structures. It is determined by the relative change in the intensity of reflected light with magnetization reversal

$$\delta = \frac{I(M)-I(-M)}{(I(M)+I(-M))/2}, \qquad (1)$$

where $I(\mathbf{M})$ is the intensity of the transmitted light in magnetized state. The magnetization direction is perpendicular to the light incidence plane. The transmission spectra were measured using the Fourier experimental setup described in Refs [10, 18]. The halogen lamp with a wide spectral range (from 0.36 to 2.5 μm) served as a light source. The radiation was collimated by a lens (focal length F is 75 mm), then it was polarized by a Glan-Taylor prism. After that, the light was focused on the sample by another lens ($F$ = 35 mm). To perform the TMOKE measurements, the sample was placed in a uniform external saturating magnetic field of 200 mT generated by an electromagnet in the direction along the gold stripes. After the sample, the light was collimated with a 20× microscope objective and detected by a spectrometer.

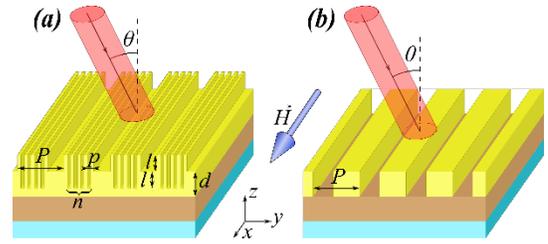

Fig. 2. Scheme of the samples and the optical measurements of the PFNL structures "A", "B" (a), and the EBL structure "E" (b).

Magnetoplasmonic nanostructures support excitation of the various optical modes, in particular, the waveguide modes and propagating SPPs. They can be revealed from the typical resonant features emerging in the transmission and MO effect spectra. In Fig. 3 the corresponding angular and wavelength-resolved spectra are given for both PFNL and EBL nanostructures.

Firstly, to observe the excitation of the propagating SPPs the phase matching condition should be met:

$$k_0 \sin\theta = k_{SPP} + mG, \qquad (2)$$

where $k_0=2\pi/\lambda$, $\lambda$ is the vacuum wavelength, $G=2\pi/P$ is a reciprocal lattice constant, $k_{SPP}$ is the SPP wavenumber, which can be approximately found by $k_{SPP} = k_0\sqrt{(\varepsilon_g\,\varepsilon_{BIG})/(\varepsilon_g+\varepsilon_{BIG})}$, $\varepsilon_g$ and $\varepsilon_{BIG}$ are dielectric constants of the metal and the magnetic layer, correspondingly, and $m$ is an integer. In our case $\varepsilon_{BIG}$ = 5.61, and $\varepsilon_g$=-20.15+1.25$i$ at a wavelength of 750 nm.

The SPP modes on the gold-magnetic layers interface are seen in the experimental transmission and the TMOKE spectra in the form of a cross with a center of symmetry at zero incidence angle (green lines in Fig. 3). At the normal incidence, the SPP modes are excited at 760 nm and 775 nm in the PFNL gratings "A" and "B", respectively. Another type of the optical modes supported by the addressed nanostructure is the waveguide modes. In this case the phase matching condition can then be described using similar equation (2), where $k_{SPP}$ is replaced with the wavenumber of the guided mode $\beta_0 = \frac{2\pi n_{\beta n}}{\lambda}$, $n_{\beta n}$ is its refractive index. In Fig. 3 one can see the waveguide modes in the spectral range from 650 nm to 750 nm that show themselves as a set of crossing lines shifted from each other. Waveguide modes are excited up to 740 nm at normal incidence for both PFNL gratings (Fig. 3(a) and (b)), and at wavelengths up to 760 nm at normal incidence for the EBL grating (Fig.3 (c)). Since the periods of the structures "A", "B" and "E" are the same, the phase matching conditions turn to be similar. Therefore, both types of the nanostructures support the same order of the SPP modes in the same spectral range.

Besides the resonance modes, waveguide and plasmonic ones, the transmission and TMOKE spectra in Fig. 3 also contain the Fabry-Perot interference pattern in the magnetic layer, which is composed of light and dark bands without an explicit dispersion dependence.

A distinctive feature of the SPP modes is their extremely high sensitivity to the properties of the metal/dielectric interface. This opens an opportunity to employ these modes as a quality learning tool for the plasmonic nanostructures (similarly to Ref. [18]).

Generally, the SPP resonance has the Fano resonance shape. The quality of the plasmonic nanostructures directly affects the parameters of the observed resonance. A comparison of the resonance frequency, width and Fano parameter for the nanostructures fabricated by two different methods one can judge the capabilities of the particular method.

The intensity of the transmitted radiation in the spectral range of the SPP resonance can be written as [27]

$$I \sim \frac{(\omega - \omega_z)^2 + \gamma_z^2}{(\omega - \omega_p)^2 + \gamma_p^2} |b_p|^2. \quad (3)$$

Here $\omega_p$ is a resonance frequency ($\lambda_p = \frac{2\pi c}{\omega_p}$ is a central wavelength of the resonance), $\gamma_p$ denotes total losses, $\omega_z = \omega_p[1 - \text{Re}(q_p)]$, $\gamma_z = \gamma_p \left[1 - \frac{\omega_p}{\gamma_p} \text{Im}(q_p)\right]$, and the corresponding wavelength is $\lambda_z = \frac{2\pi c}{\omega_z}$, $b_p$ is a parameter characterizing the amount of radiation emitted by scattering from a gratings, $a_p$ characterizes the resonant processes, $q_p = \frac{a_p}{\omega_p b_p}$ is a Fano parameter. The latter one shows what part of the field was subjected to the resonant process. Based on this, we can talk about the efficiency of the system in understanding its suitability for the excitation of modes of the structure, such as an SPP. Since the same set of modes was observed in all three structures, they can be compared. Table 1 contains the resonance parameters values found from fitting of the observed experimental spectra based on equation (3).

**Table 1. Parameters of resonances for the investigated structures at the incidence angle of 5 degrees.**

|   | $\hbar\omega_p$ (eV) | $\lambda_p$ (nm) | $\|q_p\|$ ×10⁻³ | $\hbar\gamma_p$ (eV) | $\hbar\gamma_z$ (eV) | $\hbar\omega_z$ (eV) | $\lambda_z$ (nm) |
|---|---|---|---|---|---|---|---|
| A | 1.539 ±0.001 | 806 | 3.9 ±0.5 | 0.030 ±0.001 | 0.025 ±0.005 | 1.54 ±0.01 | 805 |
| B | 1.526 ±0.003 | 813 | 2.8 ±0.8 | 0.032 ±0.003 | 0.029 ±0.006 | 1.53 ±0.02 | 811 |
| E | 1.495 ±0.003 | 829 | 67.4 ±1.3 | 0.098 ±0.002 | 0.194 ±0.008 | 1.46 ±0.02 | 847 |

The exact resonance position and shape depend on the grating geometry and are slightly different for PFNL and EBL gratings. Firstly, one can observe the change in position of $\lambda_p$ that occurs due to the fact that the SPP arises at the interface between a dielectric and not continuous, but structured metal layer.

The shape of the resonance is specified by $|q_p|$ parameter. Indeed, the efficiency of the SPP excitation as well as the Fano parameter in neatly modulated structures "A" and "B" is lower than in the etched structure "E". Thus, for "A" and "B" structures, the resonance in the transmission spectrum is distinguished by a clearly pronounced transmission minimum, and for "E" structure, by a maximum.

The quality factor of the resonance can be characterized by the resonance broadening $\hbar\gamma_p$. As can be seen from Table 1, the total losses of the PFNL structures are 3 times lower than in the etched structure. This can be clearly seen also in the TMOKE spectra in Fig. 3 (d,e,f), where the SPP resonance for "E" structure is apparently wider along the wavelengths axis.

The transmission and the TMOKE spectra for "A" and "B" structures are quite similar. The difference is bound to the resonance excitation efficiency $|q_p|$. This distinction can be seen from the TMOKE magnitude and is connected to the grooves number $n$.

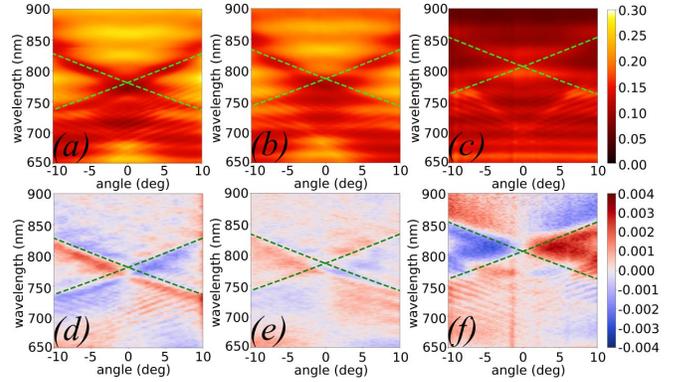

Fig. 3. Experimental transmission (a,b,c) and the TMOKE (d,e,f) spectra of "A" and "B" PFNL and "E" etched grating, accordingly. Green dashed lines correspond to the SPP resonances. The set of resonances in the range from 600 to 750 nm corresponds to the excitation of the waveguide modes in magnetic dielectric.

To analyze magneto-optical properties of the PFNL structures a rigorous coupled wave analysis was used. The best agreement of theoretical spectra with an experimental one is obtained if the scratching depth $l$ and squeezed hills height were taken the same. (Fig. 2(a)). The transmission and the TMOKE spectra dependence on $d$ were analyzed (Fig. 4). The following parameters were used: $P$ = 690 nm, $p$ = 100 nm, $n$ = 4. To exclude the Fabry-Perot and the waveguide modes, the magnetic layer was considered semi-infinite. The excitation of the SPP modes was observed in the transmission and the TMOKE for $l \geq 25$ nm.

It can be seen that the resonance position shifts upward with increasing the scratching depth $l$ (Table 2). At the same time the resonance width $\hbar\gamma_p$ tends down, so the quality factor grows. It is worth noting that $|q_p|$ grows with $l$. Therefore, the resonance excitation efficiency grows with the scratching depth. This also results in the increase of the TMOKE (Fig. 4). Therefore, varying the scratching depth and $p$ parameter, one can change the key properties of the resonance.

**Table 2. Parameters of the resonances for simulated structures depending on the depth of scratching at 5 deg incidence angles.**

| $l$ (nm) | $\lambda_p$ (nm) | $|q_p| \times 10^{-3}$ | $\hbar\gamma_p$ (eV) | $\lambda_z$ (nm) |
|---|---|---|---|---|
| 25 | 790 | 2.2±0.1 | 0.025±0.001 | 789 |
| 30 | 798 | 2.7±0.1 | 0.023±0.001 | 798 |
| 35 | 818 | 3.3±0.1 | 0.026±0.004 | 817 |
| 40 | 824 | 5.4±0.1 | 0.028±0.004 | 825 |

It's also interesting to compare the simulated and experimental spectra. The TMOKE spectrum for simulated structure with $l$ = 25 nm (Fig. 4 (a)) is qualitatively the same as for "B" structure (Fig. 3 (e)). Despite the fact that modelled structure parameters $p$ and $n$ are the same as for "B" structure, there are also similarities in spectra for "A" structure. Thus, the TMOKE spectra for simulated structures with $l$ = 30, 35 nm (Fig. 4 (b, c)) are similar to the spectrum for "A" structure (Fig. 3 (d)). In other words, an increase of the number of scratches $n$ acts as an increase of the scratch depth $l$. We can conclude that with the help of the proposed fabrication method, the result can be achieved either by varying the number of scratches or their depth. This expands the possibilities for setting a necessary configuration when fabricating structures.

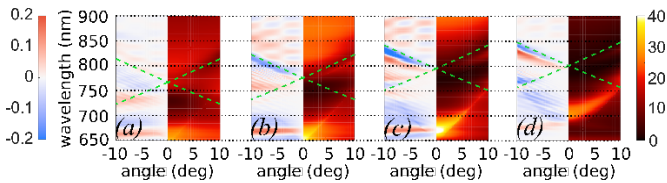

Fig. 4. Simulated TMOKE and transmission spectra of the PFNL-type structures. Different grooves depths are considered: $l$ = 25 (a), 30 (b), 35 (c), and 40 nm (d).

To sum up, the novel method of the plasmonic nanostructures fabrication was proposed. It is based on the pulse force nanolithography method. This technique provides a fine-tuning of the metal layer geometry and allows one to excite a high-quality SPP modes. Indeed, the PFNL-made nanostructures demonstrate 3-times lower total losses of the SPP resonance with respect to the grating fabricated by the electron beam lithography. The proposed method provides opportunities for the control of the SPP excitation efficiency by varying the scratching depth or the number of the scratches in a single period. The magneto-optical effect was employed to evaluate the quality of the fabricated gratings.

**Funding sources and acknowledgments.** This work was financially supported by the Ministry of Science and Higher Education of the Russian Federation, Megagrant (project No. 075-15-2019-1934). RFBR supported the theoretical study (project No. 19-02-0856) and partly supported the atomic force microscopy measurements (project No. 18-29-27020). The equipment of the "Educational and Methodical Center of Lithography and Microscopy", Lomonosov MSU was used. A.N.K. and V.I.B. are members the Interdisciplinary Scientific and Educational School of Moscow University "Photonic and Quantum technologies. Digital medicine".

**Disclosures.** The authors declare no conflicts of interest.